# Gambling on the existence of other universes

By *Arturo* SANGALLI

**Abstract**. Speculations and theories about the existence of other worlds have a long history. In recent times, the arguments have shifted away from their typical philosophical and theological character to (supposedly) become more objective thanks to their scientific underpinnings. A prime example of this is the current parallel universes or multiverse theory, which has the support of a number of famous cosmologists.

 In this article, we contend that the claim for the existence of those parallel universes, as presented in Max Tegmark's book *Our Mathematical Universe*, rests crucially on some questionable probability arguments involving infinity. As a result, a doubt is cast over the multiverse hypothesis: Is it more credible than the counterarguments based on philosophical and metaphysical considerations?

 What we call "the Universe" is in fact just one among an infinite number of other universes, so far away from ours that they are impossible to observe. In infinitely many of those "parallel" universes there are exact and almost exact living copies of you and me, and yet other universes obey different laws of physics.

 The above paragraph lists some of the astonishing consequences of the so-called multiverse theory, as presented by Max Tegmark, a physics professor at MIT, in his book *Our Mathematical Universe: My Quest for the Ultimate Nature of Reality* [T]. Although Tegmark is not alone in promoting the multiverse idea, he is one of its most visible advocates due to his press and television interviews and for contributing to numerous science documentaries.

 Critics dismiss this multiverse stuff as mere philosophical speculation since it has no empirical content, i.e. no contact with experiments, together with the fact that there is not a shred of evidence to support it nor any hint on how such evidence could come about. How then does Tegmark, a respectable scientist, defend his more-science-fiction-than-real-science story?

 The short answer is that the existence of the multiverses is a prediction of certain theories. For instance, those parallel universes, including our doppelgängers, would be a logical consequence of cosmic inflation, a theory developed in an attempt to solve some serious problems that afflict the Big Bang cosmological model.

Let's listen to Tegmark (our boldface for emphasis):

"*It feels extremely unlikely that your life turned out exactly as it did, since it required so many things to happen: Earth had to form, life had to evolve, […], your parents had to meet*, etc. *But the probability of all these outcomes happening clearly isn't zero, since it in fact happened right here in our Universe.* **And if you roll the dice enough times, even the most unlikely things are guaranteed to happen.** *With infinitely many Level I parallel universes created by inflation,* **quantum fluctuations effectively rolled the dice infinitely many times, guaranteeing with 100% certainty that your life would occur in one of them. Indeed, in infinitely many of them**, *since even a tiny fraction of an infinite number is still an infinite number*. **And an infinite space doesn't contain only exact copies of you. It contains many people that are almost like you, yet slightly different**."

To reach such stunning conclusions Tegmark appears to be applying some result from probability theory; which one, we can only guess, but his reference to dice being rolled infinitely many times points to the Borel-Cantelli lemma, a theorem frequently used concerning infinite sequences of trials. Although Tegmark never mentions it, the lemma is worth discussing if only to illustrate the question of the interpretation of probability when applied to reality—in particular if infinity is involved.

**Probability and Reality**

The following is a version of the lemma due to Émile Borel [Bo]:

*Let $T_1, T_2, …, T_n, …$ be an infinite sequence of random "trials", each of which has one of two possible outcomes: Success or Failure. Let $p_n$ be the probability of $T_n$ resulting in Success. If the infinite series $p_1 + p_2 + … + p_n + …$ converges, then the probability for Success to occur infinitely many times is equal to 0; if the series diverges, then this probability is equal to 1.*

Reflecting on his theorem, Borel cautiously warns us of the perils of applying his result outside the domain of mathematics:

"*It is easy to conceive that results such as the above can only be applied in the realm of mathematics, where we can effortlessly imagine the possibility of repeating an experiment infinitely many times.*"

But what if we ignored Borel's advice and applied his theorem to other domains such as, for example, a random sequence of letters?

The proposition below is a logical consequence of Borel's theorem:

*If symbols from the usual 26-letter alphabet plus a "blank" symbol (or space) are chosen at random to generate an infinite sequence S: $a_1, a_2, a_3, \ldots$ , then the probability for any given string of k letters (for any k) to occur infinitely many times in S is equal to 1.*

Suppose that, through some sort of mechanism or device, we would be able to actually generate such a sequence—I must admit that the "infinite" part would be a problem: there are no infinite sequences in nature, but we can imagine one through a sort of thought experiment. What would the above proposition tell us about our sequence? In particular, does it guarantee with 100% certainty that, say, the sentence "*we are by now so used to seeing reality accommodate itself to numerical rules that it is at times difficult to appreciate the astonishing fact that those rules should exist at all*" occurs an infinite number of times? Or, for that matter, that the complete works of Shakespeare will be recreated, not just once but over and over, with chance replacing the mind of the literary genius?

There is reason to doubt it. First of all, we cannot guarantee that the "random" in the proposition and the one in the generating mechanism mean the same thing—the former is theoretical; the other, well, we don't really know what it is or how to manufacture it. In fact, in the proposition, "random" simply means that every symbol has the same probability of being chosen, where "probability" is just a name for a number between 0 and 1—in this case 1/27—and therefore "the probability of event E is equal to 1" does not necessarily have as factual counterpart "event E took (or will take) place".

Theorems in probability—and a fortiori those involving the elusive notion of infinity— are mathematical results; on the other hand, assuming that for a given physical system there is a sample space satisfying the hypotheses of the theorem is an empirical claim. Confidence in predictions based on probability is no substitute for observation and verification.

The point I'm trying to make with this example is that existence in the real world cannot convincingly be established on the basis of a probabilistic result (on this, see for example [Bu]). But that is exactly what Max Tegmark appears to do to "guarantee with 100% certainty" the existence of infinitely many copies of the Earth and each of its inhabitants.

Of course, Tegmark's reference to dice being rolled is not to be taken literally. But what is then his real, serious argument?

By way of explanation, Tegmark offers the following: "*We've observed that these random-looking seed fluctuations exist, so we know that* some *mechanism [not necessarily inflation] made them*" and that this mechanism "*operated such that any*

*region could receive any possible seed fluctuations*". And he adds: "*We've measured their statistical properties using cosmic-background and galaxy maps, and their random properties are consistent with what's known to statisticians as a 'Gaussian random field'* ".

From the above, and the assumption of an infinite space and infinite matter, there would follow the property of the Level I multiverse that "everything that can happen according to the laws of physics does happen", and it happens an infinite number of times. "*This means that there are parallel universes where you never get a parking ticket, where you have a different name […] where Germany won World War II, where dinosaurs still roam Earth, and where Earth never formed in the first place.*"

But where is the evidence, either experimental or derived from some physical principle, that the random properties of these seed fluctuations would produce every possible universe—assuming that "every possible universe" is a meaningful concept?

Tegmark does mention experimental data (cosmic-background and galaxy maps, measured statistical properties, and so forth). However important these data might be, his far-reaching conclusions about the existence of parallel universes "*where Germany won World War II, where dinosaurs still roam Earth*", etc., hinge crucially on a probabilistic argument.

**Incredible luck**

Do we owe the existence of our planet and its inhabitants to a stroke of luck of cosmic proportions—to "outrageous fortune," as the title of an article in *Nature* [Br] put it?  This question is prompted by the fact that certain physical constants, such as the masses of elementary particles, the strengths of the fundamental forces, and so forth, appear to have been "fine-tuned" precisely for life on Earth to exist. Had those numerical values been ever so slightly different, terrible things would have happened—the Universe would have collapsed or atoms would never have formed, for example—preventing the formation of life. Is it just incredible luck, or did some higher entity (a deity or an advanced universe-simulating life form) design our Universe deliberately fine-tuned to allow intelligent life?

Tegmark asks this very same question and his answer is: neither. As inflation keeps eternally propagating through space, he tells us, it creates an infinite collection of Level I multiverses referred to as the Level II multiverse. Now, according to Tegmark,  "*If there are laws or constants of nature that can in principle vary from place to place, then eternal inflation will make them do so across the Level II multiverse. […] A theory where the knobs of nature take essentially all possible values will predict with 100% certainty that a habitable universe like ours exists, and since we can only live in a habitable universe, we shouldn't be surprised to find ourselves in one.*"

Put it simply: laws and fundamental constants of physics can be what they are here in our universe because they can be different in infinitely many other universes. In other words, using the die-rolling analogy, it's next to impossible to get all the constants exactly right in just one throw of the dice—this is asserted as a self-evident truth. However, if the dice were rolled infinitely many times, one should expect with "*100% certainty*", according to Tegmark, that in some cases the throw would result in a habitable universe just like ours. But where does the "100% certainty" come from? From experience or some physical principle? From a hidden assumption or postulate? We are not told.

 To sum up: it is not the existence of other, faraway universes—either as a prediction of inflation or as a mere possibility—that is hard to conceive, but the claim that the entire history of our universe, leading up to my own life, played out in exactly the same way in infinitely many of them. In the absence of experimental evidence to back it up, such possibility is much too implausible to be accepted as a consequence of inflation theory on the basis of probability arguments involving infinity. Seen in this light, Tegmark's claims appear, at best, as little more than a gamble.

**A literary digression**

 In his short story *The Immortal* [L], Jorge Luis Borges imagines a society whose members live forever. "*Taught by centuries of living*", he writes, "*the republic of immortal men had achieved a perfection of tolerance, almost of disdain. They knew that **over an infinitely long span of time, all things happen to all men*** (my emphasis). *As reward for his past and future virtues, every man merited every kindness—yet also every betrayal, as punishment for his past and future iniquities. Much as the way in games of chance heads and tails tend to even out, so cleverness and dullness cancel and correct each other. Viewed in that way, all our acts are just, though also unimportant. There are no spiritual or intellectual merits. Homer composed the Odyssey; **given infinite time**, with infinite circumstances and changes, **it is impossible that the Odyssey should not be composed at least once**.*"

 Tegmark's assertions about infinity, stated as if they were self-evident truths, are reminiscent of those found in the above passage. But then Borges' is a literary work, not a scientific one.

**Epilogue**

 It is possible that my criticism of Tegmark's multiverses due to his questionable probabilistic arguments should be unfounded, and that all those parallel universes, in which "everything that can happen according to the laws of physics does happen", really exist. If such were the case, I could take some consolation in the fact that in infinitely many of those universes the flaws in my own argument would not be discovered—not noticing reasoning errors is surely compatible with the laws of physics—and I would then be (undeservedly) praised for debunking a grandiose theory that (rightly) claimed to have elucidated the ultimate nature of reality.

**References**

[Bo] Émile Borel, "*Le jeu, la chance*", Gallimard (1941), 200-204.

[Br] Geoff Brumfiel, *Outrageous Fortune*, Nature, **493** (2006), 10-12.

[Bu] Mario Bunge, "*Evaluating Philosophies*", Boston Studies in the Philosophy of Science, **295** (2012), 157.

[L] Allen Lane, "*Borges – Collected Fiction*", Penguin Press (1954).

[T] Max Tegmark, "*Our Mathematical Universe: My Quest for the Ultimate Nature of Reality*", Alfred A. Knopf, (2014).

______________________________________________________________________

Arturo Sangalli, Sherbrooke, QC, Canada; asangall@hotmail.com

6 July, 2016.